\documentclass[12pt,preprint]{aastex}

\slugcomment{September 20, 2013; accepted to ApJ Letters}

\shorttitle{Cold dust and warm gas in NGC 4125}
\shortauthors{Wilson et al.}

\begin{document}


\title{Cold dust but warm gas in the unusual elliptical galaxy NGC 4125}


\author{C. D. Wilson\altaffilmark{1}, A. Cridland\altaffilmark{1},
  K. Foyle\altaffilmark{1}, 
T. J. Parkin\altaffilmark{1}, E. Mentuch Cooper\altaffilmark{1,2},
H. Roussel\altaffilmark{3},
M. Sauvage\altaffilmark{4},
M. W. L. Smith\altaffilmark{5},
M. Baes\altaffilmark{6},
G. Bendo\altaffilmark{7},
M. Boquien\altaffilmark{8}, A. Boselli\altaffilmark{8}, L. Ciesla\altaffilmark{8},
D. L. Clements,\altaffilmark{9}, 
A. Cooray\altaffilmark{10}, 
I. De Looze\altaffilmark{6},
M. Galametz\altaffilmark{11},
W. Gear\altaffilmark{5},
V. Lebouteiller\altaffilmark{4}, S. Madden\altaffilmark{4},
M. Pereira-Santaella\altaffilmark{12},
\and
A. R\'emy-Ruyer\altaffilmark{4}}

\altaffiltext{1}{Department of Physics \& Astronomy, McMaster University,
    Hamilton, ON L8S 4M1 Canada}
\altaffiltext{2}{Department of Astronomy, University of Texas, Austin,
TX 78712 USA}
\altaffiltext{3}{Institut d'Astrophysique de Paris, Universit\'e
  Pierre et Marie Curie, CNRS UMR 7095, F-75014 Paris, France}
\altaffiltext{4}{Laboratoire AIM, CEA/DSM-CNRS-Universit\'e Paris
  Diderot DAPNIA/Service d'Astrophysique, B\^at. 709, CEA-Saclay,
  91191 Gif-sur-Yvette Cedex, France } 
\altaffiltext{5}{School of Physics \& Astronomy, Cardiff University, The Parade, Cardiff CF24 3AA, UK }
\altaffiltext{6}{Sterrenkundig Observatorium, Universiteit Gent, Krijgslaan 281 S9, 9000 Gent, Belgium }
\altaffiltext{7}{UK ALMA Regional Centre Node, Jordell Bank Center for
  Astrophysics, School of Physics and Astronomy, University of
  Manchester, Oxford Road, Manchester M13 9PL, UK} 
\altaffiltext{8}{Aix-Marseille Université, CNRS, LAM (Laboratoire 
  d'Astrophysique de Marseille) UMR 7326, 13388 Marseille, France}
\altaffiltext{9}{Astrophysics Group, Imperial College London, Blackett Laboratory, Prince Consort Road, London SW7 2AZ, UK}
\altaffiltext{10}{Department of Physics \& Astronomy, University of California, Irvine, CA 92697, USA }
\altaffiltext{11}{Institute of Astronomy, University of Cambridge, Madingley Road, Cambridge CB3 0HA, UK}
\altaffiltext{12}{Istituto di Astrofisica e Planetologia Spaziali, INAF, Via
  Fosso del Cavaliere 100, I-00133 Roma, Italy}



\begin{abstract}
Data from the {\it Herschel Space
Observatory}\footnote{Herschel is an ESA space observatory with
science instruments provided by European-led Principal Investigator
consortia and with important participation from NASA}  have revealed an unusual elliptical galaxy, NGC 4125, which has
strong and extended submillimeter emission from cold dust but only very strict
upper limits to its CO and HI emission. 
Depending on the dust emissivity, 
the total dust mass is
$2-5 \times 10^6$ M$_\odot$. While the neutral gas-to-dust mass ratio  is
extremely low ($< 12-30$), including the ionized gas traced by [CII] emission
raises this limit to $< 39-100$.
The dust emission follows a similar
$r^{1/4}$ profile to the stellar light and the dust to stellar mass ratio
is towards the high end of what is found in
nearby elliptical galaxies. 
We suggest that NGC 4125 is
currently in an unusual phase where evolved  stars produced in
a merger-triggered burst of star formation are  pumping large
amounts of gas and dust into the interstellar medium. In this
scenario, the low neutral gas-to-dust mass ratio is explained by the gas
being heated to temperatures $\ge 10^4$~K faster than the dust is
evaporated. 
If galaxies like NGC 4125, where the far-infrared emission does not trace
 neutral gas in the usual manner,
are common at higher redshift, this could have significant
implications for our understanding of high redshift galaxies and galaxy
evolution.
\end{abstract}





\section{Introduction}

Early-type galaxies (ellipticals and lenticulars) are typically
dominated by an old stellar population and contain at most a small
fraction of their total mass in cold gas and dust. Recent surveys of
nearby galaxies have detected CO emission in
5\% of the elliptical galaxies in the ATLAS$^{\rm 3D}$ sample 
\citep{y11} and submillimeter
emission from cold dust in 24\% of the elliptical galaxies in the
Herschel Reference Survey \citep{s12}.
%\citep{s12,c12}.
Unfortunately there are relatively few early-type galaxies that
have been observed and detected in both dust and gas tracers. 
\citet{l08} measured gas-to-dust mass ratios of $\sim 300-400$ in a sample
of 7 elliptical galaxies.
\citet{s12} measure gas-to-dust mass ratios
for 8 S0 galaxies that are consistent with
or higher than the value of 100-150 typically found in spiral galaxies.
% secondary removal
%Higher resolution CO observations of 12 early-type galaxies
%(including two ellipticals) show primarily central concentrations of
%molecular gas with a good spatial correlation with ionized gas tracers
%\citep{c11}. 

\object{NGC 4125} is a luminous E6(pec) galaxy at a distance of 23.9 Mpc
\citep{t01}, which underwent a merger and a 
burst of star formation 6-8 Gyr ago \citep{s92,p10}.
Its 60 $\mu$m IRAS luminosity of $10^{8.7}$ L$_\odot$ places
it at the high luminosity end of the SAURON sample
\citep{c07}. 
%\citet{c07} find that the correlation of molecular gas mass with
%far-infrared luminosity for early-type galaxies in the SAURON sample
%is similar to that of spiral galaxies.
However, NGC 4125 is undetected in either HI or CO
\citep{w10}; tight upper limits correspond to $M_{HI} < 2.4\times 10^7$
M$_\odot$ (2$\sigma$, 8.7$^\prime$ beam) and $M_{H_2} < 1.9\times
10^7$ M$_\odot$ (2$\sigma$, 21$^{\prime\prime}$ beam) when using a CO-to-H$_2$
conversion factor of $2\times 10^{20}$ H$_2$ cm$^{-2}$ (K km
s$^{-1}$)$^{-1}$ \citep{s88}. 
It has significant X-ray emission from hot gas, with at most a few
percent contribution by an AGN \citep{b11}. 
%Among a sample of 30 normal
%(non-cD) early-type galaxies observed with Chandra, NGC 4125 lies in
%the top decade of both $L_K$ and $L_X$ \citep{b11}. 
%% log(LK) = 11.3 
%NGC 4125 is one of five elliptical galaxies included in the SINGS
%sample \citep{k03} and has the reddest $NUV-r$ colors of the early
%type galaxies \citep{d07}.
%
% tertiary removal
%Its mid-infrared spectrum shows polycyclic aromatic hydrocarbon
%as well as 
%H$_2$ emission lines \citep{s07,d09} and 
 %would therefore likely be classified as a Case 3 (``star formation'')
 %or Case 2 (``post star formation'') elliptical by \citet{p11}.
 %
%At mid-infrared wavelengths, NGC 4125 stands out from most elliptical
%galaxies in having a high ratio of 70 $\mu$m to 24 $\mu$m luminosity
%\citep{t09}. 
It was selected as the ``normal'' elliptical galaxy for
 the Very Nearby Galaxies Survey with the {\it Herschel
Space Observatory} (PI: C. D. Wilson). However, as we describe in this
Letter, the interstellar medium (ISM) of NGC 4125 appears to be anything
but normal. 

\section{{\it Herschel} observations}\label{obs}

%% In a manner similar to \objectname authors can provide links to dataset
%% hosted at participating data centers via the \dataset{} command.  The
%% second curly bracket argument is printed in the text while the first
%% parentheses argument serves as the valid data set identifier.  Large
%% lists of data set are best provided in a table (see Table 3 for an example).
%% Valid data set identifiers should be obtained from the data center that
%% is currently hosting the data.
%%
%% Note that AASTeX interprets everything between the curly braces in the 
%% macro as regular text, so any special characters, e.g. "#" or "_," must be 
%% preceded by a backslash. Otherwise, you will get a LaTeX error when you 
%% compile your manuscript.  Special characters do not 
%% need to be escaped in the optional, square-bracket argument.

NGC 4125 was observed in scan map mode over a $9^\prime\times 9^\prime$ area
with the {\it Herschel Space Observatory }
%at 70 and 160 $\mu$m 
with the Photodetector Array Camera and Spectrometer
\citep[PACS,][OBSIDs 1242188202,
1242188203]{pog10} and the Spectral and Photometric Imaging Receiver
\citep[SPIRE,][OBSID 1242199156]{g10}.
The data processing and
calibration for the SPIRE data are described in \citet{b12}. 
For the
PACS photometry, the level-1 data were downloaded from the Herschel
Science Archive using HIPE 9.0 and the latest calibration
(PACS\_CAL\_41\_0) and  
mapped using {\sc Scanamorphos} v21 \citep{r13}. 
NGC 4125 was observed in the 
[CII] 158 $\mu$m, [NII] 122 $\mu$m, and [OI] 63 $\mu$m  emission lines
(OBSIDS 1342222063, 1342222064, 1342222065) 
using PACS in chop/nod mode with a single 
pointing covering $47\times 47^{\prime\prime}$. 
The maps
were processed as described in \citet{p13} using HIPE and the PACSman
package \citep{l12}. 
The 250 $\mu$m image of NGC 4125 is shown overlaid on the K-band image
from 2MASS \citep{j03} in Figure~\ref{fig-Kcomp}.

The total continuum flux of NGC 4125 was measured %from these maps
using an elliptical aperture 
$ 216^{\prime\prime} \times 118^{\prime\prime}$ in diameter (62\% of
the $D_{25}$ optical size) with
position angle 82$^o$; the sky background was measured in an
elliptical annulus with inner and outer diameters of 
216$^{\prime\prime}$ and 360$^{\prime\prime}$. 
%This diameter is 62\% of the optical $D_{25}$
%size of NGC 4125.
%For the PACS maps, the total flux within an 
%18$^{\prime\prime}$ (2.1 kpc) circular aperture was also measured for 
%comparison with the peak SPIRE 250 $\mu$m flux. 
%The flux was corrected for the mean sky
%level, which was measured
%in a similarly-shaped elliptical annulus with inner diameter
%216$^{\prime\prime}$ and outer diameter 360$^{\prime\prime}$. The
%global and sky apertures were selected to maximize the total flux and
%minimize the uncertainties while keeping the same size aperture for
%all wavelengths.
For the SPIRE data, in calculating the total flux, we adopt beam areas
appropriate for emission going as $\nu^4$
of 432, 766 and 1573 square arcseconds 
at 250, 350,
and 500 $\mu$m, respectively, and 
apply the multiplicative color corrections 
appropriate for extended sources %with emission going as $\nu^4$ 
\citep{som11}.
For the PACS data, we adopt the color corrections appropriate for a 20
K blackbody \citep{m11}.
For both sets of maps, 
the uncertainties are calculated as the quadrature sum of the
instrumental uncertainty in each pixel and the uncertainty due to the
sky confusion.
%The confusion uncertainty in the peak flux is given by the 
%standard deviation of the sky annulus, $\sigma_{sky}$,  and the
%confusion uncertainty in the total 
%flux is given by $\sigma_{sky} \sqrt{N_{beams}}$, where $N_{beams}$ is
%the number of SPIRE beams in the aperture. 
Calibration uncertainties
of 7\% for SPIRE \citep{som11} and 5\%  for PACS
\citep{pom11} are added in quadrature to these
measurement uncertainties in Table~\ref{tbl-phot}.  

%The PACS 70 and 160 $\mu$m fluxes were measured using the same apertures
%for the source and the sky. In addition, the total flux within an 
%18$^{\prime\prime}$ (2.1 kpc) circular aperture was measured for 
%comparison with the peak SPIRE 250 $\mu$m flux. 
%The uncertainties are calculated as the quadruature sum of the
%instrumental uncertainty in each pixel and the uncertainty due to the
%sky confusion.
%For both the peak and the total
%flux, the confusion uncertainty was calculated
%as $\sigma_{sky} \sqrt{N_{pix}}$, where $N_{pix}$ is the number of
%pixels within the aperture.
%The calibration uncertainties of 5\% 
%\citep{pom11} 
%are added in quadrature to the
% measurement uncertainties. 

\begin{figure}
\epsscale{0.8}
\includegraphics[angle=0]{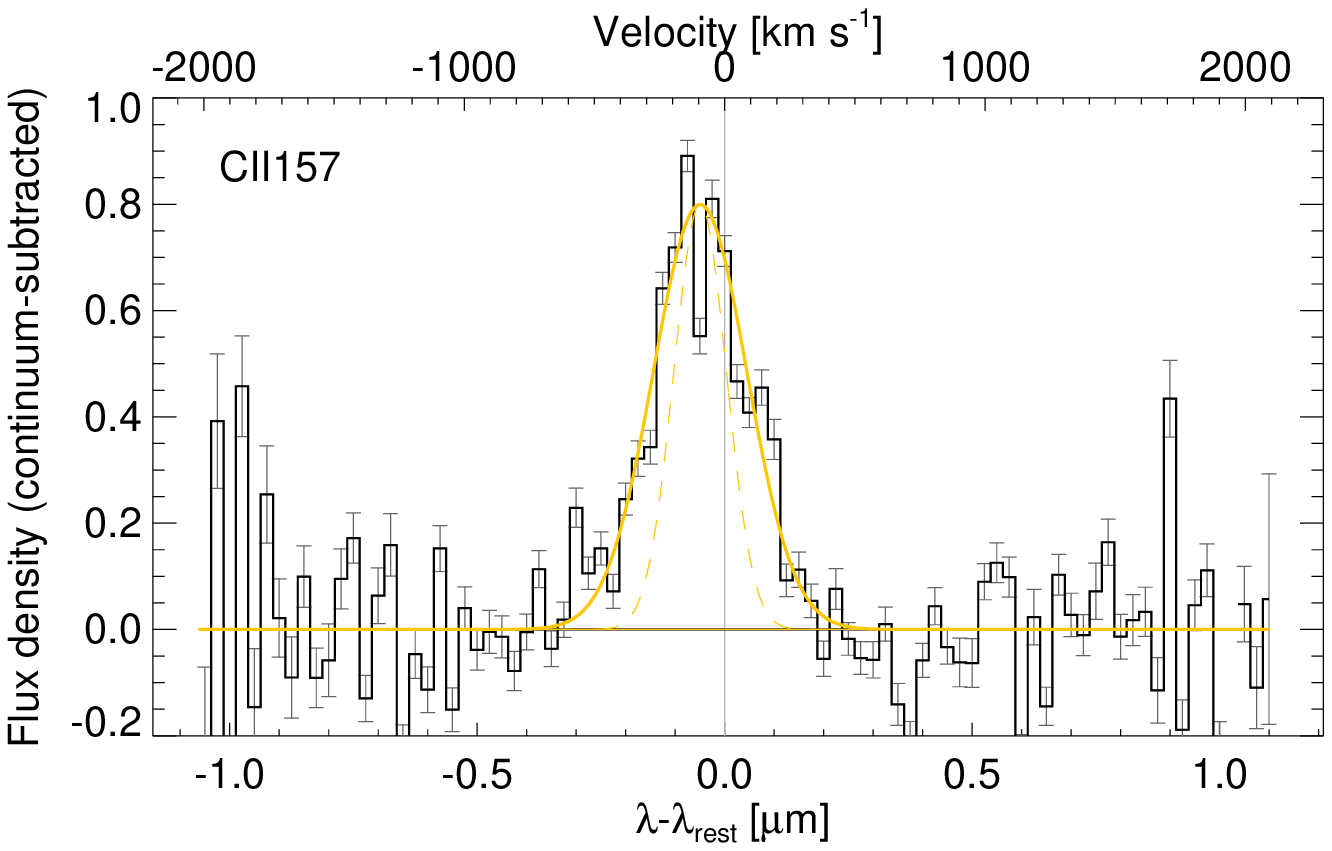}
\includegraphics[angle=0]{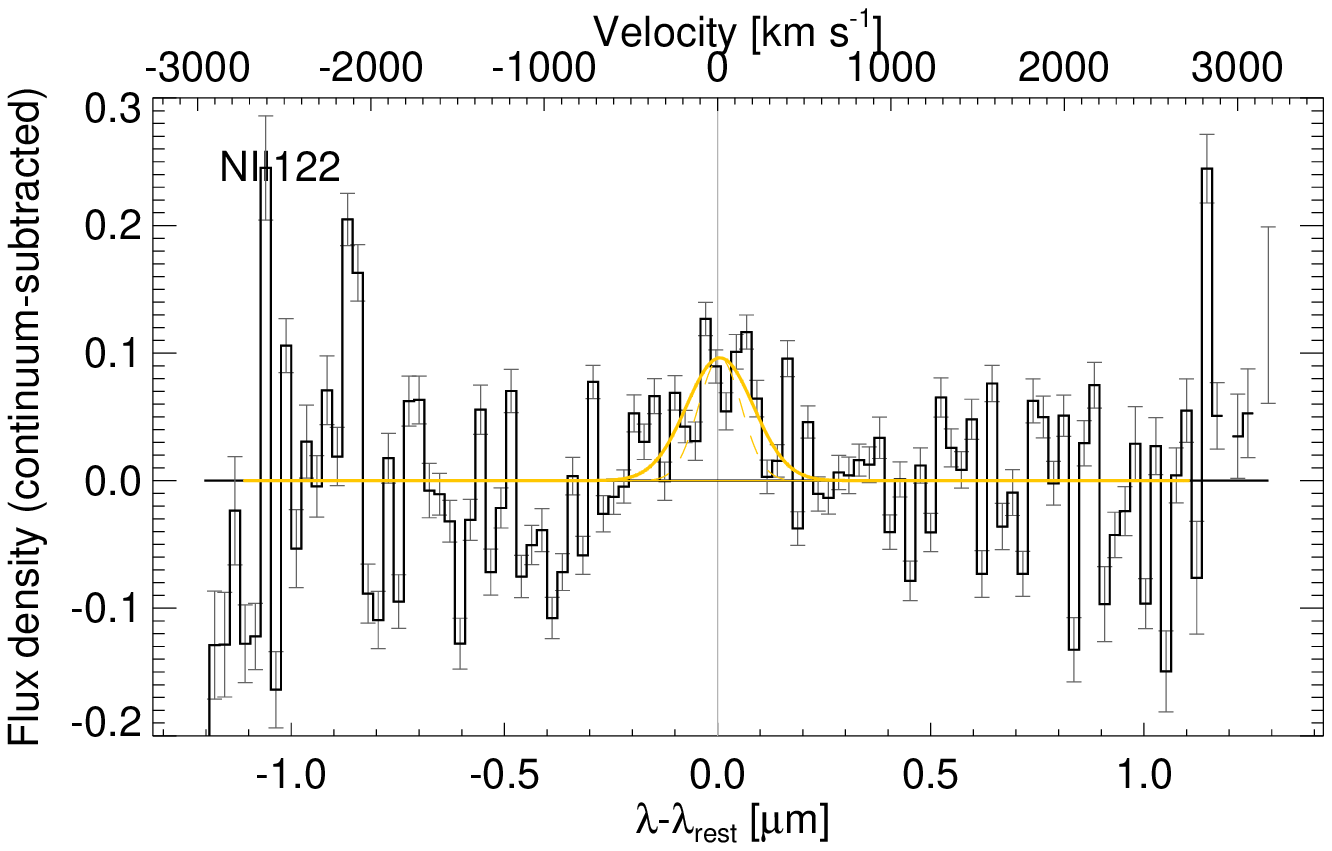}
\caption{(top) Central [CII] 158 $\mu$m spectrum of NGC 4125. The yellow solid line
  shows the fit to the data; the dashed yellow line indicates the
  intrinsic velocity resolution of the instrument. (bottom) Central
  [NII] 122 $\mu$m spectrum of NGC 4125.
\label{fig-PACS}}
\end{figure}

We have detected weak [CII] and [NII] emission
(Figure~\ref{fig-PACS}).
The peak [CII] flux in an 18$^{\prime\prime}$ diameter aperture is
$(4.34 \pm 0.07) \times 10^{-17}$ W
m$^{-2}$ and the total flux summed in a 36$^{\prime\prime}$ diameter
aperture is
$(9.63 \pm 0.14)\times 10^{-17}$ W m$^{-2}$. For [NII], the peak flux is
$(9.0 \pm 0.4) \times 10^{-18}$ W m$^{-2}$ and the total flux is
$(2.19 \pm 0.08)\times 10^{-17}$ W m$^{-2}$. The [OI] 63 $\mu$m line
is not detected, with a 2$\sigma$ upper limit to the total flux of
$7.5\times 10^{-18}$ W m$^{-2}$.
Calibration uncertainties are 30\% for these spectroscopic data.

\section{The gas-to-dust mass ratio in NGC 4125}\label{gd}

\begin{figure}
\epsscale{0.8}
\includegraphics[scale=.51]{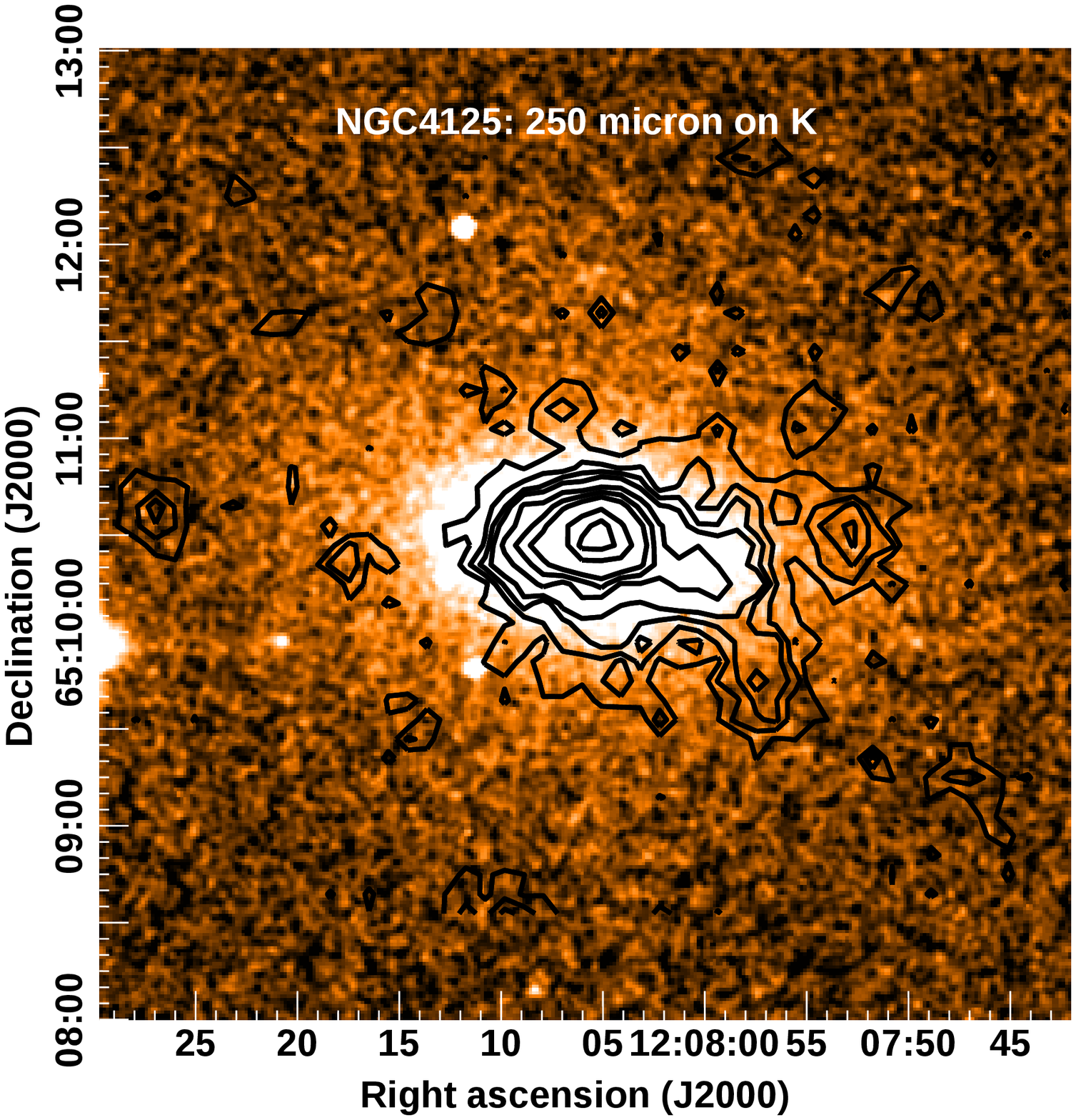}
\includegraphics[angle=-90,scale=.65]{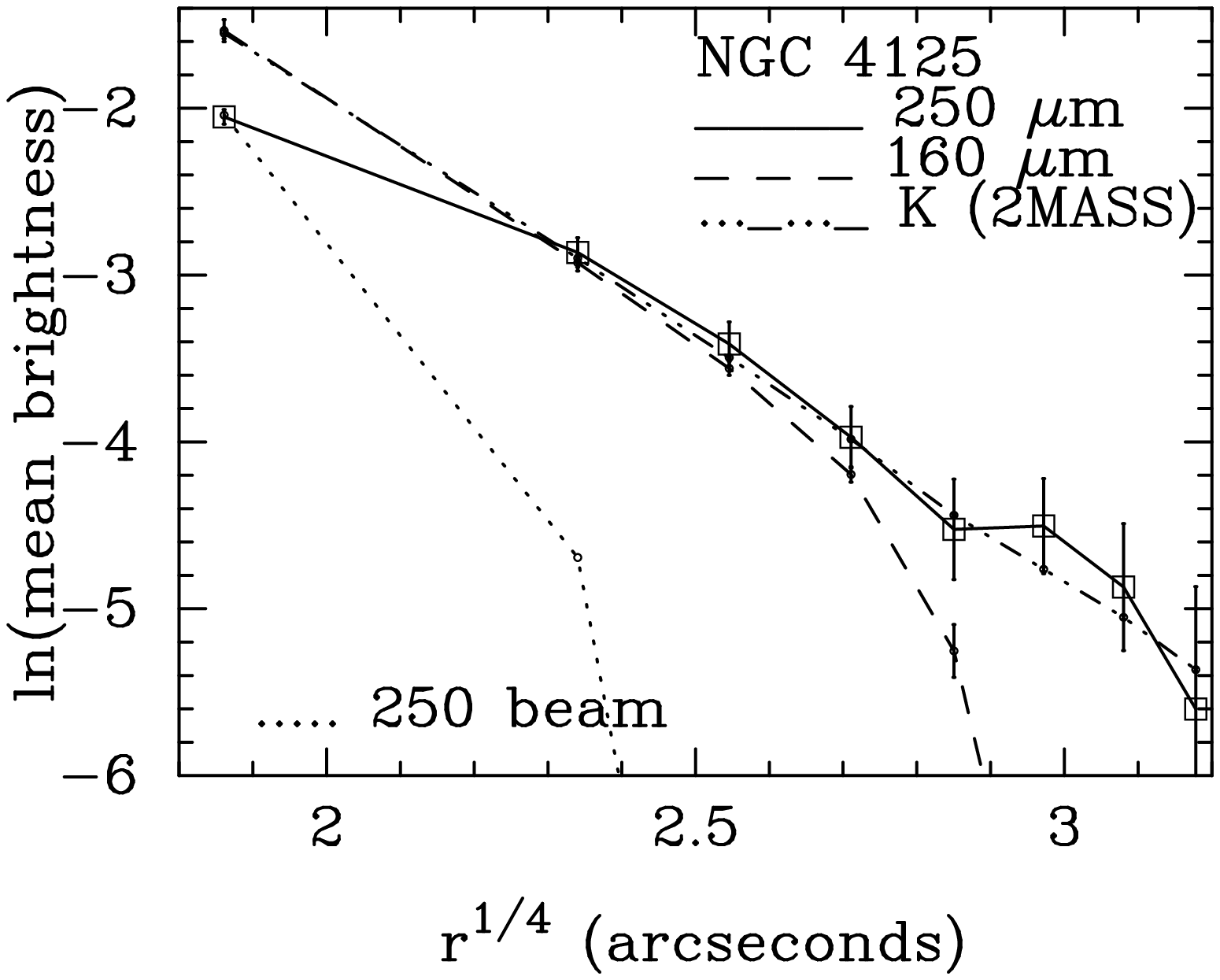}
\caption{(Top) 250 $\mu$m contours overlaid on the $K$-band image from 2MASS.
Contour levels    are 10, 20, 30, 40, 60, 80, 100, 150, and 200 mJy
beam$^{-1}$.
The two point-like sources to the west could be background sources and 
may be responsible for the slight excess in the 500 $\mu$m emission
(Figure~\ref{figBB}).
(bottom) Radial profiles of the 250 $\mu$m (solid),
160 $\mu$m (dashed) and K-band
brightness (dot-dashed) from 2MASS. The dotted line
shows the 250 $\mu$m beam profiled measured in the same annuli
as the radial profile.
\label{fig-Kcomp}}
\end{figure}

One unusual aspect of the dust emission in NGC 4125 is that the
emission is spatially extended and follows the general shape of the
stellar emission (Figure~\ref{fig-Kcomp}). While previous optical extinction
maps have shown a central dust lane aligned with the major axis
\citep{g94}, the {\it Herschel} images trace dust over a significantly
larger area and the dust emission
even follows the $r^{1/4}$ profile that is a common feature of
ellipticals. 
%This radial profile is most easily traced in the 250
%$\mu$m image, which has the best combination of sensitivity and
%angular resolution, but the galaxy is clearly spatially extended in
%all the continuum maps.

\begin{figure}[htb]
\centering
\includegraphics[scale=0.8]{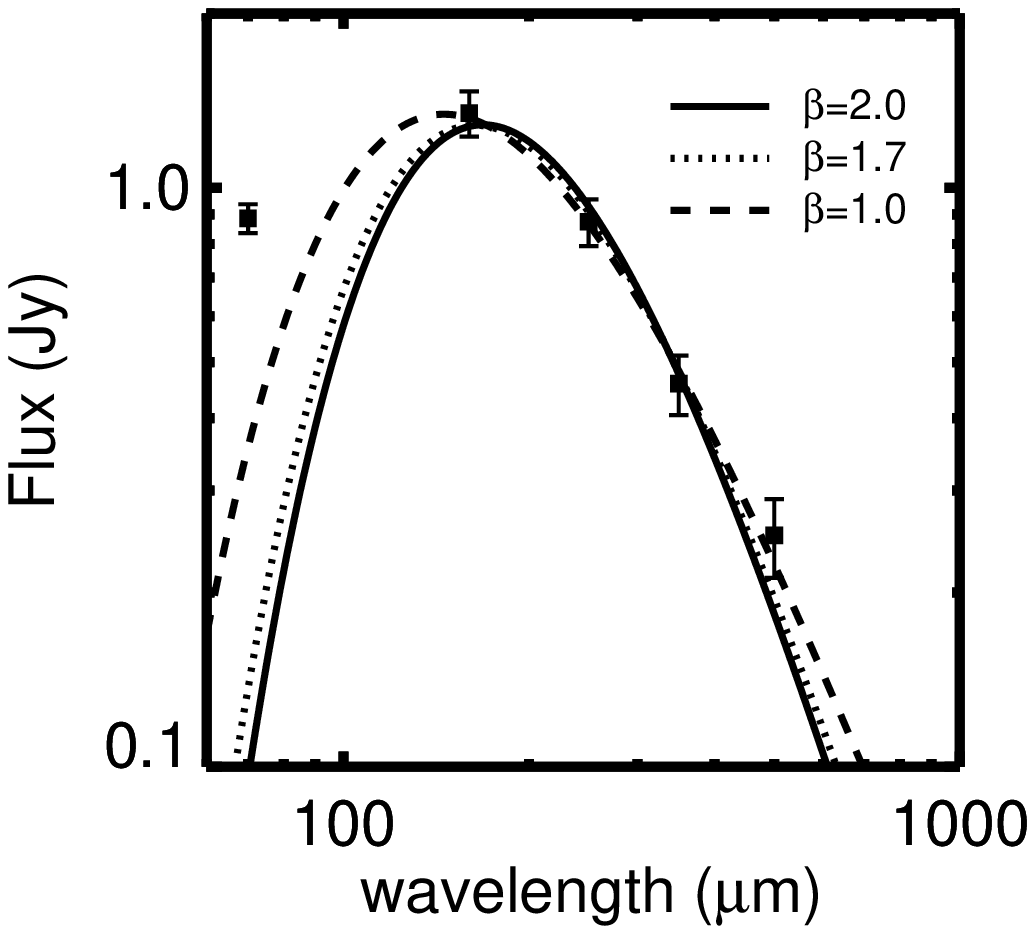}
\includegraphics[scale=0.8]{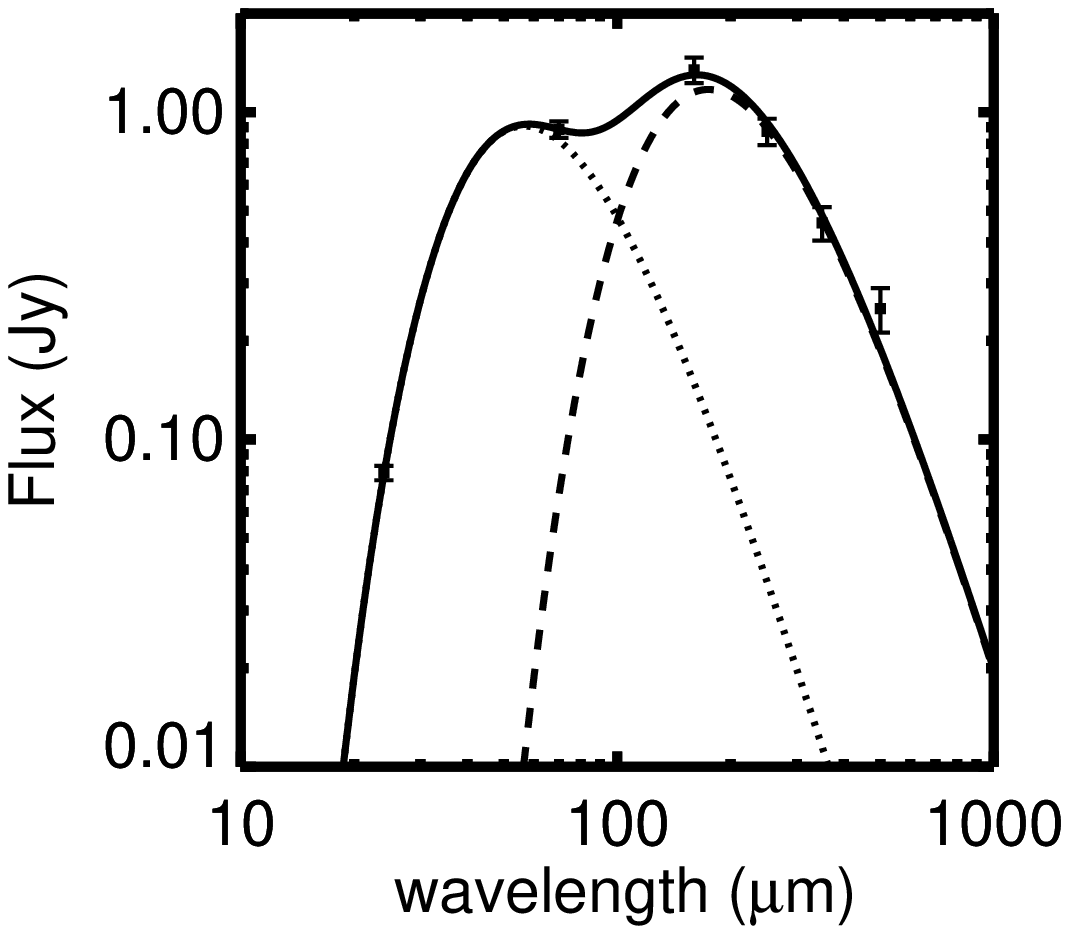}
\caption{(top) The global fluxes for NGC 4125 measured from the {\it Herschel}
  maps with three modified blackbody fits  to the data from
160 to 500 $\mu$m with $\beta=2$ (solid), 
$\beta=1.7$ (dotted) and $\beta=1$ (dashed) overlaid. 
(bottom) A two temperature component fit to the data from
24 to 500 $\mu$m with $\beta=2$.
\label{figBB}}
\end{figure}

%The relatively weak 1.4 GHz radio continuum emission of 1.9 mJy
%\citep{b07} is not a significant source of contamination in the {\it
%  Herschel} maps.
Fitting the global dust emission from 160 to 500 $\mu$m with a
modified blackbody with dust opacity going as $\nu^2$ gives an average
dust temperature of $17.2 \pm 0.8$ K (Figure~\ref{figBB}). 
The dust mass can be
calculated as
$M_{dust} = 1449 D_{Mpc}^2 S_{250} (\exp (57.58/T)-1)/\kappa_{250}$ M$_\odot$,
where $D_{Mpc}$ is the distance to the galaxy in Mpc, $T$ is the dust
temperature in K, 
$S_{250}$ is the 250 $\mu$m flux in Jy, and $\kappa_{250}$
is the dust emissivity at 250 $\mu$m in cm$^{2}$ g$^{-1}$. As our
default, we assume 
a standard graphite plus silicate dust emissivity $\kappa_{250} =
3.98$ cm$^2$ g$^{-1}$ 
\citep{d03}, which corresponds to $\beta \simeq 2$ and has also been
used for previous analyses of elliptical galaxies \citep{p12,s12}.
We obtain a global dust mass for NGC 4125
of $5.0^{+1.5}_{-1.1} \times 10^6$ M$_\odot$ (Table~\ref{tbl-prop}) and a mass in the peak
resolution element of $(1.2\pm 0.2)\times 10^6$ M$_\odot$.  
Using the same distance and emissivity, our
dust mass is a factor of 4 larger than that of
\citet{k11}, who obtained a warmer dust temperature due to a lack of
fluxes longward of 160 $\mu$m.
% secondary removal
%This dust mass
%is also significantly larger than the mean dust mass of $\sim 3\times 10^5$
%M$_\odot$ measured for the elliptical galaxies detected by
%\citet{s12} that have similar K-band luminosities to NGC 4125.

The dust mass is rather sensitive to our
assumptions about the grain properties. If we instead adopt the
mixture of amorphous carbon and silicates 
explored for the Large Magellanic Cloud by \citet{g11}, for which
$\beta=1.7$ and 
$\kappa_{250} = 7.49$ cm$^2$ g$^{-1}$,
we obtain a dust temperature of $19.0\pm 1.0$ K  and a total dust mass
of
$(1.9^{+0.6}_{-0.4}) \times 10^6$ M$_\odot$. 
%This dust mass is a
%factor of 2.6 times smaller 
%than for our more standard graphite plus silicate dust model. 
Performing the fit with $\beta$ allowed to vary between 1 and 2.5
returns a fit with $\beta=1$ and a temperature of $25\pm 2$
K. Combining this temperature with $\kappa_{250}=3.98$ cm$^2$ g$^{-1}$
gives similar
dust masses to the amorphous carbon model of \citet{g11}.
We also note, however, that single temperature fits to spiral galaxies
usually yield smaller dust masses than more realistic multiple
temperature models \citep{d12}.
Fitting the data with two components with $\beta=2$ 
(Figure~\ref{figBB}) gives 
temperatures of 17 and 51 K 
and a cold dust mass of $5.4 \times 10^6$ M$_\odot$.
% new starting here
Although the 500 $\mu$m flux lies somewhat above the fits,
the excess emission is likely due to two faint ($\sim 25$ mJy) point-like
sources to the west of NGC 4125 (Figure~\ref{fig-Kcomp}). These sources
contribute 20\% of the total flux at 500 $\mu$m but only 10\% at
250 $\mu$m and may be background galaxies.
% new ending here

% xxx
%Finally,
%for both the $\beta=2$ and $\beta=1.7$ solutions, the 500 $\mu$m flux
%gives masses that are 30-40\% higher than the masses obtained from the
%250 or 350 $\mu$m fluxes. Thus, there may be a small excess of
%500 $\mu$m emission in NGC 4125, although not at the level that is found
%in some low-metallicity dwarf galaxies \citep{g12}.
%Any 500 $\mu$m excess would drive the blackbody fits towards lower
%values of $\beta$.  

We estimate a stellar mass of $2.4 \times 10^{11}$ M$_\odot$ from the
3.6 $\mu$m flux in 
\citet{d07} and the median mass-to-light ratio of
1.00 $M_\odot$ $L_\odot^{-1}$ from \citet{fb11}. The  dust to
stellar mass ratio of 
$2.1 \times 10^{-5}$ is towards the high end of the values seen for
nearby elliptical galaxies \citep{s12,dSA13}.
Comparing the global dust mass with
the sum of the HI and H$_2$ upper limits 
\citep{w10} and including a factor of 1.36 to account for helium
\citep{w88}
gives a global gas-to-dust
ratio of $< 12$. Because the areas covered by the three different
measurements (HI, CO, dust) are very different, another relevant comparison is
between the peak dust mass (18$^{\prime\prime}$ beam) and the CO upper
limit (21$^{\prime\prime}$ beam). Comparing these two values gives a
limit on the gas-to-dust mass ratio of $< 22$.
These $2\sigma$ upper limits are factors of  7-12
smaller than the typical gas-to-dust mass ratio of 150 seen in our own and
other galaxies. Thus, there appears to be a deficit of neutral gas to go
along with the cold dust in NGC 4125.

We can also estimate the gas mass associated with the observed
[CII] emission.  We assume that the [CII] 
emission comes from the ionized gas phase ($T\sim 8000$ K) via
collisional excitation with electrons.  
We note that the [NII]/[CII] ratio is consistent with a 
low-density ionized medium \citep{o06}. 
%\citet{k11} use mid-infrared
%[SII] line ratios to obtain an upper
%limit to the electron density of $n_e < 30$ cm$^{-3}$.
Assuming the electron density, $n_e$, is much less 
than the critical density $n_{crit} \sim 35$ cm$^{-3}$ \citep{m93}, 
% xx MRP: At 8000K I obtain n_crit = 43 cm-3 = 2.3e-6 s-1/(4.8e-8 cm3 s-1
%     *(0.8)^-0.5)  using the atomic parameters from the LAMDA database
the column density is given by
$N_H = 2.13\times 10^{20} { I_{C+} \over X_{C+}} {n_{crit} \over {
    n_e}}$
with $I_{C+}$ in units of erg s$^{-1}$ cm$^{-2}$ sr$^{-1}$. 
Assuming a C$^+$
abundance relative to hydrogen of $X_{C+} = 3\times 10^{-4}$, 
the mass corresponding to the total flux in the map is
$1.09 \times 10^7 /n_e$ M$_\odot$.
Solving for the case where the ionized medium uniformly fills the 
 18$^{\prime\prime}$ (2.1 kpc)
radius volume gives a lower limit to $n_e$ of 0.11 cm$^{-3}$ and an
 ionized hydrogen mass  of  $\le 1.0 \times
10^8$ M$_\odot$. Combining this mass estimate with our dust mass and
again accounting for helium with a factor of 1.36 gives
a total (ionized plus neutral) gas-to-dust mass ratio of $< 39$ 
 for $\beta=2$ or $< 100$ for $\beta=1.7$. 
% secondary removal
%If we adopt the amorphous carbon ($\beta=1.7$) emissivity, the global
%gas-to-dust mass ratio is $< 80$. 
%
% tertiary removal
%The minimum gas-to-dust mass ratio for solar metallicity
%gas containing 1.7\% metals by mass would be 60 if none of the heavy elements
%reside in the gas phase \citep{g11}. 
This analysis suggests that a significant fraction of the gas
in NGC 4125 is warm ionized gas that produces 
[CII] emission  rather than cold neutral gas.

%Finally, we have calculated $L_{CII}/L_{FIR}$ for comparison with
%other recent studies. Although NGC 4125 is not a starburst, Seyfert,
%or luminous infrared galaxy, it lies nicely on the relationship
%between $L_{CII}/L_{FIR}$ and $L_{FIR}/M_{H_2}$ shown by
%\citet{gc11}. Its $L_{CII}/L_{FIR}$ ratio appears similar to those of other
%metal-rich, mostly spiral galaxies \citep{b08}.

\section{Discussion}

\subsection{An internal origin for the dust in NGC 4125?}

The dusty ISM of NGC 4125 has some unusual properties.
The dust has a similar radial profile 
to the stellar light out to at least 7 kpc (1$^\prime$) radius 
(Figure~\ref{fig-Kcomp}). 
Its total dust mass is among the highest seen in elliptical
galaxies \citep{s12}
and is only a factor of 3 times smaller than the
dust mass in the elliptical galaxy Centaurus A \citep{p12}. 
However,
unlike Cen A, where the  neutral gas-to-dust mass ratio is 100 
and the gas traces a clearly rotating disk \citep{p12}, in
NGC 4125 
the corresponding gas is not  neutral but in a warmer ionized or even x-ray emitting
phase.
These unusual properties suggest
an internal origin for the ISM of NGC 4125.

The ratio $M_{dust}/M_{*}$ is towards the high end of the
range seen in ellipticals, as is the total stellar mass $M_*$
\citep{s12}. 
 The optical colors of NGC 4125 suggest that it underwent a
burst of star formation 6-8 Gyr ago \citep{s92}. Long-slit
spectroscopy combined with simple stellar population models indicate
super-solar metallicity and also point to a recent dissipational
merger event and stellar ages as young as 8 Gyr \citep{p10}.
Many of  these younger stars would now be in the red giant, carbon star,
or asymptotic giant branch (AGB) phase when stars shed large
quantities of gas and dust. 
The material that is lost from these stars 
has a normal gas-to-dust mass ratio.
% and so cannot by itself explain
% the low gas-to-dust mass ratio in NGC 4125. 
However, if the dust from the
star survives for a longer time \citep[i.e. 50 Myr, ][]{c10} 
than the gas survives in the  neutral phase, the
observed {\it neutral} gas to dust mass ratio could be reduced, because the
majority of the gas
would be in an ionized or X-ray emitting phase. 
% secondary removal
%Although there have been no detailed models for the mass of the X-ray
%emitting gas in NGC 4125, its hot gas is more luminous (although
%25\% cooler) than the hot gas in NGC 720, for which \citet{h11}
%estimate a total mass in hot gas of $4.3\times 10^{10}$ M$_\odot$.
%
% tertiary removal
%Thus, although NGC 4125 shows no evidence for a significant cold
%neutral medium, there is likely a massive gas 
%reservoir present in the ionized and X-ray emitting gas.

Assuming a dust lifetime of 50 Myr, the stars in NGC 4125 would need 
to produce dust at a rate of $\sim 0.1$ M$_\odot$ yr$^{-1}$ to
explain the observed dust mass.
This rate is much higher than the typical rate seen in passive
elliptical galaxies \citep{c10}.
\citet{f06} derived total (gas plus dust) mass loss rates from a few $\times 10^{-8}$
to a few $\times 10^{-3}$ M$_\odot$ yr$^{-1}$ for a collection of AGB
stars and proto-planetary and planetary nebulae
in our own Galaxy, with a median mass loss rate of 
$10^{-5}$ M$_\odot$ yr$^{-1}$. \citet{bo12} find that extreme AGB stars
are responsible for $\sim 90\%$ of the dust produced by evolved stars
in the SMC.
Assuming a normal gas-to-dust mass
ratio of 150, NGC 4125 would need to contain roughly $2\times 10^6$
such stars (0.001\% of its total stellar mass) to produce the required
dust. Given the total mass of 
NGC 4125, such a population of stars seems quite reasonable.

 In contrast, if the dust in NGC 4125 was acquired via an interaction with another galaxy, the lack of neutral gas suggests that the interaction must have been extremely recent. We would need to be viewing NGC 4125 at the
precise time when most of the gas has been ionized but significant cold dust remains. 
%If the dust in NGC 4125 is contained in a central disk (similar to that in Cen A), its average %surface density is a factor of 10 lower than the bright regions of Cen A \citep{p12}. 
It is hard to understand how such a large mass of dust could survive for much more than 50 Myr without the high column densities which would  shield both the dust and the neutral gas.  

% paragraph: secondary removal 
%On the other hand, if the dust in NGC 4125 was acquired externally
%through mergers and 
%interactions, the amount of material required is quite large
%($\sim 10^9$ M$_\odot$ of gas with a normal gas-to-dust mass ratio).
%While Cen A is clearly an example of an elliptical galaxy that has
%accreted a comparable quantity of gas which has had time to settle
%into a central disk, it is hard to explain the lack of {\it cold} gas
%in NGC 4125 under an accretion scenario. If NGC 4125 acquired its gas
%and dust externally, we would need to be viewing the galaxy at a 
%precise time when much of the gas has been ionized and is no longer
%cold, while the dust remains in a cold state. Given that the expected
%lifetime of a dust grain in contact with the hot X-ray medium is only
%of order 1 Myr \citep{c10},
%an external origin for the dust in NGC 4125 seems quite unlikely.

\subsection{How unique is NGC 4125?}

One question which arises is why we have not discovered more galaxies
like NGC 4125. First, such galaxies may be intrinsically rare 
 in the local Universe.
The Herschel Reference Survey \citep{b10} 
contains all elliptical galaxies in a portion of the northern sky with
$K \le 8.7$ 
mag and $15 < D < 25$ 
Mpc; NGC 4125 meets the distance and luminosity criteria but falls
just outside the 
upper declination limit of the survey. 
 Adding NGC 4125 to
the 18 elliptical galaxies
 from \citet{s12}, we can estimate that the rate for NGC 4125-like
galaxies may be $\sim 
5$\%.  However,  not all the galaxies in the Herschel Reference
Survey have atomic and molecular gas measurements and so some could be
similar to NGC 4125.
Second, NGC 4125 is a luminous galaxy, which
makes it easier to detect a given mass fraction of dust. 
For internal processes, a higher
stellar mass would correlate with a higher dust mass.
The presence of a younger stellar population than in
passively evolving ellipticals could also serve to increase the dust
to stellar mass ratio. Given
the sensitivities of typical {\it Herschel} maps, a similar galaxy which was
1/3 as massive would appear as a point source at 250 $\mu$m at the
same distance as NGC 4125,
while a galaxy 1/10 as massive could not be distinguished easily from
the {\it Herschel} confusion limit. 
Finally, it is hard to know how critical the exact age and mass
fraction of the younger stellar population  might be in producing this
particular configuration of dust without  neutral gas.
% tertiary removal
%Stellar evolution models \citep{sch92} combined with a standard 
%initial mass function 
%suggest that a population of stars from 1 to 3
%M$_\odot$ sheds similar aggregate amounts of gas and dust per year, as the
%lower amount of mass shed by a lower mass star is balanced  by the
%fact that lower mass stars are more numerous. 
With stellar lifetimes
of 1.4 Gyr for a 2 M$_\odot$ star and 5.5 Gyr for a 1.25 M$_\odot$ star
\citep{sch92}, the exact age of the young
population may not be a critical factor, as long as a young enough
population forms a sufficient fraction of the stellar mass.

Understanding the frequency and origin of galaxies like NGC 4125 may have
important implications for understanding red galaxies at larger
redshifts. 
A galaxy like NGC 4125 would be rather difficult to interpret if it
was observed at much larger distances. Its relatively
strong submillimeter emission could be interpreted 
as indicating the presence of a significant  cold ISM
\citep{e10},
while its rest-frame 70 $\mu$m emission could
be taken as 
an indicator of on-going star formation. % \citep{c05}.
For example, using the star formation rate calibration of \citet{li10},
the 70 $\mu$m flux in the central 18$^{\prime\prime}$  could
be interpreted
as a star formation rate of 0.15 M$_\odot$ yr $^{-1}$. For a typical
gas depletion time of 2 Gyr \citep{leroy08}, such a star formation
rate would imply 
an H$_2$ mass of $3\times 10^8$ M$_\odot$, a mass which is clearly
ruled out by the existing CO upper limits.
At some point in the past, the progenitors of elliptical galaxies must
have contained a significant  ISM in order to form
their current population of stars. 
 If the ISM in NGC 4125 has an internal origin
linked to the age of its
most recent 
star formation episode, 
then we might expect galaxies like NGC 4125 to be more
common at larger redshifts, when the overall stellar population of
elliptical galaxies was, of necessity, younger.

Recently, observations of moderate redshift galaxies have
identified optically red galaxies which contain significant infrared
emission. 
% secondary removal
%For example, \citet{s08}  report 24 $\mu$m
%detections of galaxies in eight massive galaxy clusters with redshifts
%from 0.02 to 0.83; some of these galaxies lie in the cluster red
%sequence. Although in this case they may not be precise analogs of NGC
%4125, since their morphology is described as late-type, similar
%processes could be at work in early-type spirals or lenticular
%galaxies. 
\citet{r12} and \citet{a13} have studied large samples of
early-type (elliptical and lenticular) galaxies from the H-ATLAS
\citep{e10b} survey. While these galaxies typically have larger dust masses
and lower stellar masses than NGC 4125, it would be interesting to
follow up on the handful of massive galaxies with red $NUV-r$ colors.
\citet{d08}
have identified a sample of
dust-obscured galaxies with  high Spitzer mid-infrared to optical flux ratios.
The $BzK$ colors of some of these galaxies show they are clearly 
forming stars \citep{p08}, but many are not detected in $B$ and
so could instead be passive, post-star formation galaxies.
{\it Herschel} surveys have also identified a number of 250 $\mu$m sources
with red optical colors \citep{s11}. While \citet{d11} conclude that
most of these are highly obscured galaxies, 
some of them have $NUV-r$ colors that are similar to NGC 4125.
If galaxies like NGC 4125, where the far-infrared emission is not
related to star formation and  neutral gas in the usual manner,
become more common at higher redshift, this would have significant
implications for our understanding of high redshift galaxies and galaxy
evolution.

\acknowledgments

 We thank the referee for comments that helped improve this paper.
This research of CDW is funded by the Canadian Space Agency
and the Natural Sciences and Engineering Research Council of Canada.
MPS has been funded by the Agenzia Spaziale
Italiana (ASI) under contract I/005/11/0. PACS has been developed by a
consortium of institutes led by MPE (Germany) and including UVIE
(Austria); KU Leuven, CSL, IMEC (Belgium); CEA, LAM (France); MPIA
(Germany); INAF-IFSI/OAA/OAP/OAT, LENS, SISSA (Italy); IAC
(Spain). This development has been supported by the funding agencies
BMVIT (Austria), ESA-PRODEX (Belgium), CEA/CNES (France), DLR
(Germany), ASI/INAF (Italy) and CICYT/MCYT (Spain). SPIRE has been
developed by a consortium of institutes led by Cardiff University (UK)
and including Univ. Lethbridge (Canada); NAOC (China); CEA, LAM
(France); IFSI, Univ. Padua (Italy); IAC (Spain); Stockholm
Observatory (Sweden); Imperial College London, RAL, UCL-MSSL, UKATC,
Univ. Sussex (UK); and Caltech, JPL, NHSC, Univ. Colorado (USA). This
development has been supported by national funding agencies: CSA
(Canada); NAOC (China); CEA, CNES, CNRS (France); ASI (Italy); MCINN
(Spain); SNSB (Sweden); STFC (UK); and NASA (USA). 
%HIPE is a joint
%development by the Herschel Science Ground Segment Consortium,
%consisting of ESA, the NASA Herschel Science Center, and the HIFI,
%PACS and SPIRE consortia. 
This research has made use of the NASA/IPAC
Extragalactic Database (NED) which is operated by the Jet Propulsion
Laboratory, California Institute of Technology, under contract with
the National Aeronautics and Space Administration.


%% in parentheses, after the keyword, but they will not be verified.

{\it Facilities:} \facility{Herschel}

\clearpage

%% the file, uncomment the \plotone, \plottwo, and
%% If you need a layout that cannot be achieved with \plotone or
%% \plottwo, you can invoke the graphicx package directly with the
%% \includegraphics command or use \plotfiddle. For more information,
%% please see the tutorial on "Using Electronic Art with AASTeX" in the
%% documentation section at the AASTeX Web site,
%% http://www.journals.uchicago.edu/AAS/AASTeX.
%%
%% The examples below also include sample markup for submission of
%% supplemental electronic materials. As always, be sure to check
%% the instructions to authors for the journal you are submitting to
%% for specific submissions guidelines as they vary from

%% This example uses \plotone to include an EPS file scaled to
%% 80% of its natural size with \epsscale. Its caption
%% has been written to indicate that additional figure parts will be
%% available in the electronic journal.

%\begin{figure}
%\epsscale{1.30}
%\plotone{ngc4125_250microns.eps}
%\caption{250 $\mu$m image of the elliptical galaxy NGC 4125. Contour levels
%    are 10, 20, 30, 40, 60, 80, 100, 150, and 200 mJy beam$^{-1}$.\label{fig1}}
%\end{figure}

\clearpage

\clearpage

\clearpage

\clearpage

%% Here we use \plottwo to present two versions of the same figure,
%% one in black and white for print the other in RGB color
%% for online presentation. Note that the caption indicates
%% that a color version of the figure will be available online.
%%

%\begin{figure}
%\plottwo{f2.eps}{f2_color.eps}
%\caption{A panel taken from Figure 2 of \citet{rudnick03}. 
%See the electronic edition of the Journal for a color version 
%of this figure.\label{fig2}}
%\end{figure}

%% Tables should be submitted one per page, so put a \clearpage before

%% Two options are available to the author for producing tables:  the
%% deluxetable environment provided by the AASTeX package or the LaTeX
%% table environment.  Use of deluxetable is preferred.
%%

%% Three table samples follow, two marked up in the deluxetable environment,
%% one marked up as a LaTeX table.

%% In this first example, note that the \tabletypesize{}
%% command has been used to reduce the font size of the table.
%% We also use the \rotate command to rotate the table to
%% landscape orientation since it is very wide even at the
%% Note also that the \label command needs to be placed

%% This table also includes a table comment indicating that the full
%% version will be available in machine-readable format in the electronic

\clearpage

\begin{table}
\begin{center}
\caption{Infrared fluxes for NGC 4125\label{tbl-phot}}
\begin{tabular}{cccc}
\tableline\tableline
Wavelength & Resolution & Peak flux\tablenotemark{a} & Total flux \\
($\mu$m) & (FWHM, $^{\prime\prime}$) & (mJy) & (mJy) \\
\tableline
24 & 6 & ... & $79\pm 4$\tablenotemark{b} \\
70 & 5.76 & $370\pm 19$ & $886 \pm 51$ \\ 
160 & 12.13 & $415\pm 22$ & $1347\pm 121$ \\ 
250 & 18.2 & $211 \pm 8$ & $874\pm 81$  \\
350 & 24.5 & $114\pm 14$ & $459 \pm 54$  \\
500 & 36.0 & $81\pm 13$ & $251\pm 39$  \\
\end{tabular}
\tablenotetext{a}{Peak fluxes with SPIRE are from maps with the native SPIRE 
resolution and have units of mJy beam$^{-1}$. Peak fluxes with PACS are 
measured in an
18$^{\prime\prime}$ circular aperture and have units of mJy.}
\tablenotetext{b}{Flux from
  \citet{d07}. }
% secondary removal
%The measured fluxes before this correction using an
%  aperture of size similar to 
%  the one adopted here are  72, 957, and 1351 mJy at 24, 70, and 160
%  $\mu$m, respectively.}
%\tablecomments{Fluxes include measurement and calibration
%  uncertainties in quadrature. Color corrections for Herschel data as
%  described in \S\ref{obs}.}
\end{center}
\end{table}
%
% took beam sizes from Tara's paper
% don't include line fluxes in same table?
%
%158\tablenotemark{b} & $\sim$12 & $(4.3 \pm 0.1)\times 10^{-17}$ & $
%(9.6 \pm 0.1) \times 10^{-17}$  & This paper \\
%122\tablenotemark{c} & $\sim$10 & $(9.0 \pm 0.4)\times 10^{-18}$ & $
%(2.2 \pm 0.1) \times 10^{-17}$  & This paper \\
%% Any table notes must follow the \end{tabular} command.

\clearpage

\begin{table}
\begin{center}
\caption{Global interstellar medium properties of NGC 4125\label{tbl-prop}}
\begin{tabular}{lcl}
\tableline\tableline
Property & Value & Notes \\
\tableline
%%$\log L_X$ & $4.9\times 10^{40}$ & erg s$^{-1}$ & \citet{b11} \\
%
% secondary removal
%$\log L_K$ & 11.3  L$_\odot$ & \citet{b11} \\
%Distance & 23.9  Mpc & \citet{t01} \\
%
%Age & 6-8  Gyr & \citet{s92,p10} \\
%Size ($D_{25}$) & 5.8$^\prime$ & \citet{deV91} \\
%%70/24 $\mu$m flux ratio & 5.0 & ... & \citet{t09} \\
%1.365 GHz flux & 1.9 mJy & \citet{b07} \\
$M_{dust}$ & $5.0^{+1.5}_{-1.1} \times 10^6$ M$_\odot$\tablenotemark{a}  & $\beta=2$, graphite + silicate grains \\
$M_{\rm HI}/M_{dust}$ & $<5$ & 2$\sigma$ upper limit\\
$M_{\rm HI+H_2}/M_{dust}$ & $<12$ & 2$\sigma$ upper limit\\
$M_{\rm HII+HI+H_2}/M_{dust}$ & $<39$ & Assumes uniform
ionized medium (\S\ref{gd}) \\
$M_*$ & $2.4 \times 10^{11}$ M$_\odot$ & ...  \\
$M_{dust}/M_*$ & $2.1 \times 10^{-5}$ & ... \\
% xx might remove these if short of space?
%$\log L_{FIR}$ & $1.2 \times 10^9$ L$_\odot$ & Using updated
%fluxes from \citet{k89}\\
%$\log L_{CII}$ & $1.7 \times 10^6$ L$_\odot$ & This paper \\
%$\log L_{CII}/L_{FIR}$ & $1.4 \times 10^{-3}$ & This paper\\
%$\log L_{FIR}/M_{H_2}$ & $> 65$ & This paper\\
\tableline
\end{tabular}
\tablenotetext{a}{If we adopt the dust emissivity appropriate for 
amorphous carbon plus silicate grains, the dust mass becomes
$1.9\times 10^6$ M$_\odot$; see text.}
\end{center}
\end{table}

\end{document}